\documentclass[aps,pra,amssymb,floats,twocolumn,superscriptaddress]{revtex4}

\usepackage{array}
\usepackage{graphicx}   
\usepackage{verbatim}   
\usepackage{color}      
\usepackage{braket}
\usepackage{amsmath}
\usepackage[]{graphicx}
\usepackage{parskip}
\usepackage{setspace}
\usepackage{amssymb}
\usepackage{float}
\usepackage{physics}
\usepackage{MnSymbol}
\usepackage{hyperref}
\usepackage{qcircuit}

\newcommand{\Renyi}[0]{R\'{e}nyi~}
\newcommand{\mysection}[1]{{ \it #1 --}}

\begin{document}

\title{Identification of symmetry-protected topological states on noisy
	quantum computers}

\author{Daniel Azses}
\affiliation{Department of Physics, Bar-Ilan University, Ramat Gan
	5290002, Israel}
\affiliation{Center for Quantum Entanglement Science and Technology,
	Bar-Ilan University, Ramat Gan 5290002, Israel}

\author{Rafael Haenel}
\affiliation{Department of Physics and Astronomy,  University of 
British
	Columbia, Vancouver, BC V6T 1Z1, Canada}
\affiliation{Stewart Blusson Quantum Matter Institute,  University of
	British Columbia, Vancouver, BC V6T 1Z4, Canada}

\author{Yehuda Naveh}
\affiliation{IBM Research - Haifa, Haifa University Campus, Mount 
Carmel,
	Haifa 31905, Israel}

\author{Robert Raussendorf}
\affiliation{Department of Physics and Astronomy,  University of 
British
	Columbia, Vancouver, BC V6T 1Z1, Canada}
\affiliation{Stewart Blusson Quantum Matter Institute,  University of 
	British Columbia, Vancouver, BC V6T 1Z4, Canada}

\author{Eran Sela}
\affiliation{School of Physics and Astronomy, Tel Aviv University, Tel
	Aviv 6997801, Israel}

\author{Emanuele G. Dalla Torre}
\affiliation{Department of Physics, Bar-Ilan University, Ramat Gan 
5290002,
	Israel}
\affiliation{Center for Quantum Entanglement Science and Technology,
	Bar-Ilan University, Ramat Gan 5290002, Israel}

\begin{abstract}
Identifying topological properties is a major challenge because, by definition, topological states do not have a local order parameter. While a generic solution to this challenge is not available yet, a broad class of topological states, namely symmetry-protected topological (SPT) states, can be identified by distinctive degeneracies in their entanglement spectrum. Here, we propose and realize two complementary protocols to probe these degeneracies based on, respectively, symmetry-resolved entanglement entropies and measurement-based computational algorithms. The two protocols link quantum information processing to the classification of SPT phases of matter. They invoke the creation of a cluster state, and are implemented on an IBM quantum computer. The experimental findings are compared to noisy simulations, allowing us to study the stability of topological states to perturbations and noise.
\end{abstract}

\maketitle

One of the most important achievements in modern physics is the 
discovery
and classification of topological phases of matter. Topological states 
do
not break any local symmetry and, hence, are robust against local
perturbations. In the context of quantum computation, this protection 
can
be used to perform quantum protocols that are robust to local noise
sources. The downside of this protection is that local probes are
insufficient to identify  topological states. Hence, even if one is 
able to
create a topological state, demonstrating its topological character 
can be
very challenging.

\begin{figure}[b]
	\vspace{-0.5cm}
	\includegraphics[width=0.95\linewidth]{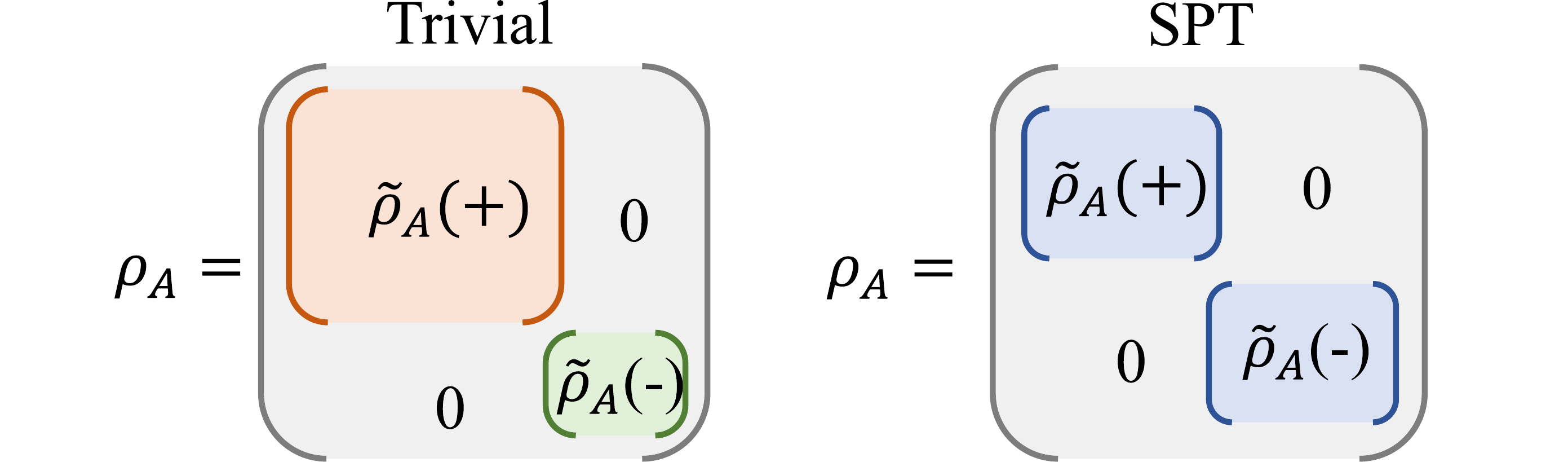}
	\vspace{-0.5cm}
	\caption{Schematic distinction between trivial 
	states and SPT ones.}
	\label{fig:schematic}
\end{figure}

In this work, we address this question for a specific class of topological states, 
known as symmetry protected topological (SPT) states. SPT phases can be identified by inspecting their entanglement spectrum
(ES), i.e., the set of eigenvalues of the reduced density matrix of a
subsystem, $\rho_A$. In particular, for ground states of one 
dimensional
(1D) SPT phases the ES is formed by degenerate pairs (or multiplets), 
while
in topologically trivial states there is no protected
degeneracy~\cite{pollmann2010entanglement,fidkowski2010entanglement,chen2013symmetry}\footnote{
The
degeneracy of the ES of an SPT state is exact only in the 
thermodynamic
limit where both the system size and the size of the subsystem 
tend to
infinity. For finite size systems, the degeneracy is exponentially
suppressed as the ratio between the (sub)system size and the 
correlation
length of the state.}. 
A simple explanation for the existence of ES degeneracies is offered 
by the
symmetry-resolved structure of
$\rho_A$~~\cite{laflorencie2014spin,goldstein2018symmetry}. Consider a 
SPT
phase protected by a unitary symmetry $G = G_A \times G_B$, where 
$G_A$ and
$G_B$ act on subsystems $A$ and $B$, respectively. Because $G$ commutes
with the Hamiltonian, the ground state of the SPT phase, $|\psi_{\rm
	gs}\rangle$, is an eigenstate of the symmetry operator $G$. When 
	performing
a partial trace $\rho_A={\rm Tr}_B[|\psi_{\rm 
gs}\rangle\langle\psi_{\rm
	gs}|]$, the conservation of $G$ guarantees that $\rho_A$ is block 
	diagonal
in $G_A$, see Fig.~\ref{fig:schematic}. One can then define
symmetry-resolved reduced density matrices as $\tilde{\rho}_A = \Pi_A
\rho_A \Pi_A$, where $\Pi_A$ projects a state on a specific symmetry
sector. 
For simple SPTs, like the Haldane phase of integer spins or Kitaev 
chains,
it was found~\cite{cornfeld2019entanglement,fraenkel2019symmetry} 
that  
$\tilde{\rho}_A$ that belong to different sectors are identical, 
leading to
a degenerate ES \footnote{A general relationship between symmetry 
sectors 
	$\tilde{\rho}_A$ in arbitrary SPTs can be derived using cohomology
	theory~\cite{Inpreparation}.}.

A related property of SPT phases is the possibility to use their ground
states as resources for measurement-based quantum computation (MBQC), where 
the process of computation is driven by local measurements  \cite{miyake2010quantum,doherty2009identifying,else2012symmetry}. A paradigmatic example of MBQC is offered by the quantum-wire protocol, whose goal is to transfer quantum information between the two edges of a one-dimensional chain. In this protocol, the input state is implemented in the protected edge state of the SPT phase and measurements are used to progressively reduce the size of the chain and transfer the information to the opposite edge. Ref.~\cite{else2012symmetry} 
established that the quantum-wire-protocol is a uniform property of all 
ground states belonging to a given SPT phase of 1D spin chains. This result 
was subsequently extended to include measurement-based quantum gates in 1D 
SPT phases \cite{miller2015resource,raussendorf2017symmetry} and finally to 
universal MBQC in 2D SPT phases \cite{raussendorf2019computationally,%
	devakul2018universal,%
	stephen2019subsystem,daniel2019computational}.

Here, we propose to use symmetry-resolved density matrices and MBQC
protocols to identify the SPT properties of a quantum state. First, we
develop and implement a new quantum protocol that accesses each 
symmetry
sector individually. The equivalence of the different sectors helps us
identify SPT states and distinguish them from trivial ones. Next, we 
extend
the MBQC wire protocol \cite{else2012symmetry} to
include local perturbations. We demonstrate that the protocol can be
disturbed only by perturbations that break the symmetry and make the 
state
trivial, hence providing a complementary method to identify SPT states.

\begin{figure}[t]	
	\includegraphics[width=1\linewidth]{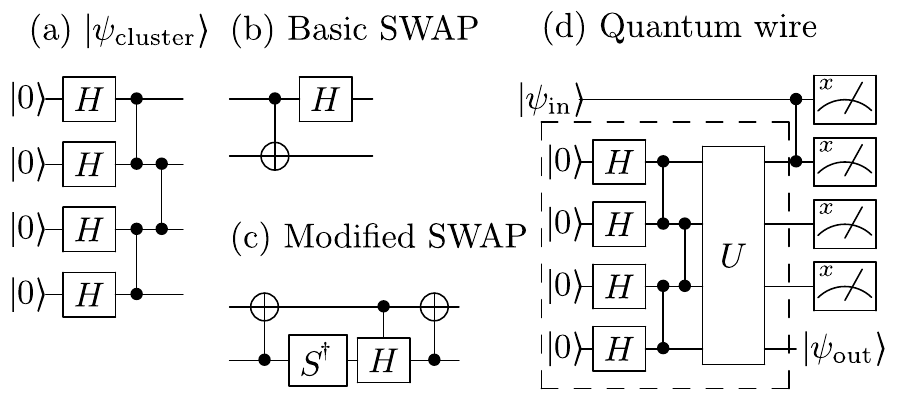}
	\caption{Building blocks of the quantum circuits used in this 
	article: (a)
		Preparation of the cluster state $|\psi_{\rm cluster}\rangle$. 
		(b) Basic
		SWAP test, which takes the singlet to $|11\rangle$, and the 
		triplets to a
		mixture of $|00\rangle,|01\rangle,|10\rangle$, reproduced from
		Ref.~\cite{garcia2013swap,cincio2018learning}.
		(c) Modified SWAP test, which identifies all four eigenvectors 
		of $(Z_i
		\otimes I) \rm{SWAP}$. This gate is used to compute 
		symmetry-resolved
		purities. (d) MBQC teleportation 
		algorithm, using the
		state $U|\psi_{\rm cluster}\rangle$ as a resource. See also 
		section F of
		the Supplemental Materials for the full quantum circuits.}
	\label{fig:circuits}
\end{figure}

\mysection{Cluster state} Having in mind the physical realization of our algorithm using qubits, we focus on the 1D cluster Ising 
Hamiltonian
\begin{align} H_{\rm cluster}= - \sum_i h_i = -\sum_{i} 
Z_{i-1}X_{i}Z_{i+1},\label{eq:Hcluster}
\end{align} where $\{X,Y,Z\}$ are Pauli matrices and $h_i$ are 
referred to as stabilizers 
\cite{gottesman1997stabilizer,briegel2001persistent,keating2004random,%
kopp2005criticality,son2011quantum,doherty2009identifying,%
smacchia2011statistical,niu2012majorana,degottardi2013majorana,%
lahtinen2015realizing,ohta2016topological,lee2016string,friedman2017phases}.
Its ground state, also known as the 1D cluster state $|\psi_{\rm 
 cluster}\rangle$, is a topological state protected by the 
$\mathbb{Z}_2\times \mathbb{Z}_2$ symmetry associated with the conservation 
of $P_{\rm odd} = \prod_i h_{2i+1} =\prod_i X_{2i+1}$ and $P_{\rm even} = 
\prod_i h_{2i} = \prod_i X_{2i}$.
These operators correspond to parities on the sublattices of odd and even 
sites, respectively. For periodic boundary conditions, the reduced density 
matrix $\rho_A$ of the cluster state has 4 identical eigenvalues $\lambda=1/4$,
one for each sector of the $\mathbb{Z}_2\times \mathbb{Z}_2$ symmetry, see
section D of the Supplemental Materials. 

The Hamiltonian $H_{\rm cluster}$ can be obtained from a trivial
Hamiltonian $H_{\rm trivial}=-\sum_i X_i$ by the transformation 
$X_i \to Z_{i-1}X_iZ_{i+1}$ and $Z_i \to Z_i$ \footnote{Incidentally, we note 
that by applying a similar transformation  $X_i \to Z_{i-1}X_iZ_{i+1}$ 
and $Y_i \to Z_i$ (with additional phase factors) multiple times, one obtains a 
series of topological states of increasing complexity 
\cite{friedman2017phases,Inpreparation}.}. Similarly, $|\psi_{\rm cluster}\rangle$ can be prepared in two steps \cite{choo2018measurement,smith2019crossing}: (i) Hadamard gates that bring the system to the ground state of  $H_{\rm trivial}$, $|\psi_{\rm trivial}\rangle = |+++...\rangle$; (ii) controlled-Z gates that realize the transformation, see Fig.~\ref{fig:circuits}(a) 
\cite{lindner2009proposal,economou2010optically,schwartz2016deterministic,%
buterakos2017deterministic,	pichler2017universal}. 
If the first and last qubits are not linked, see Fig.~\ref{fig:circuits}(a), one 
obtains a system with open boundary conditions. In this case, the first 
and last terms of the corresponding Hamiltonian, see Eq.~(\ref{eq:Hcluster}), 
become $h_1=X_1 Z_2$ and $h_L=Z_{L-1}X_L$ and the state conserves the total 
parity $P = (-1)^L \prod_{i=1}^L h_i = Y_1X_2X_3\dots X_{L-1}Y_{L}$.

\begin{figure}[t]
	\centering
	\includegraphics[width=1\linewidth]{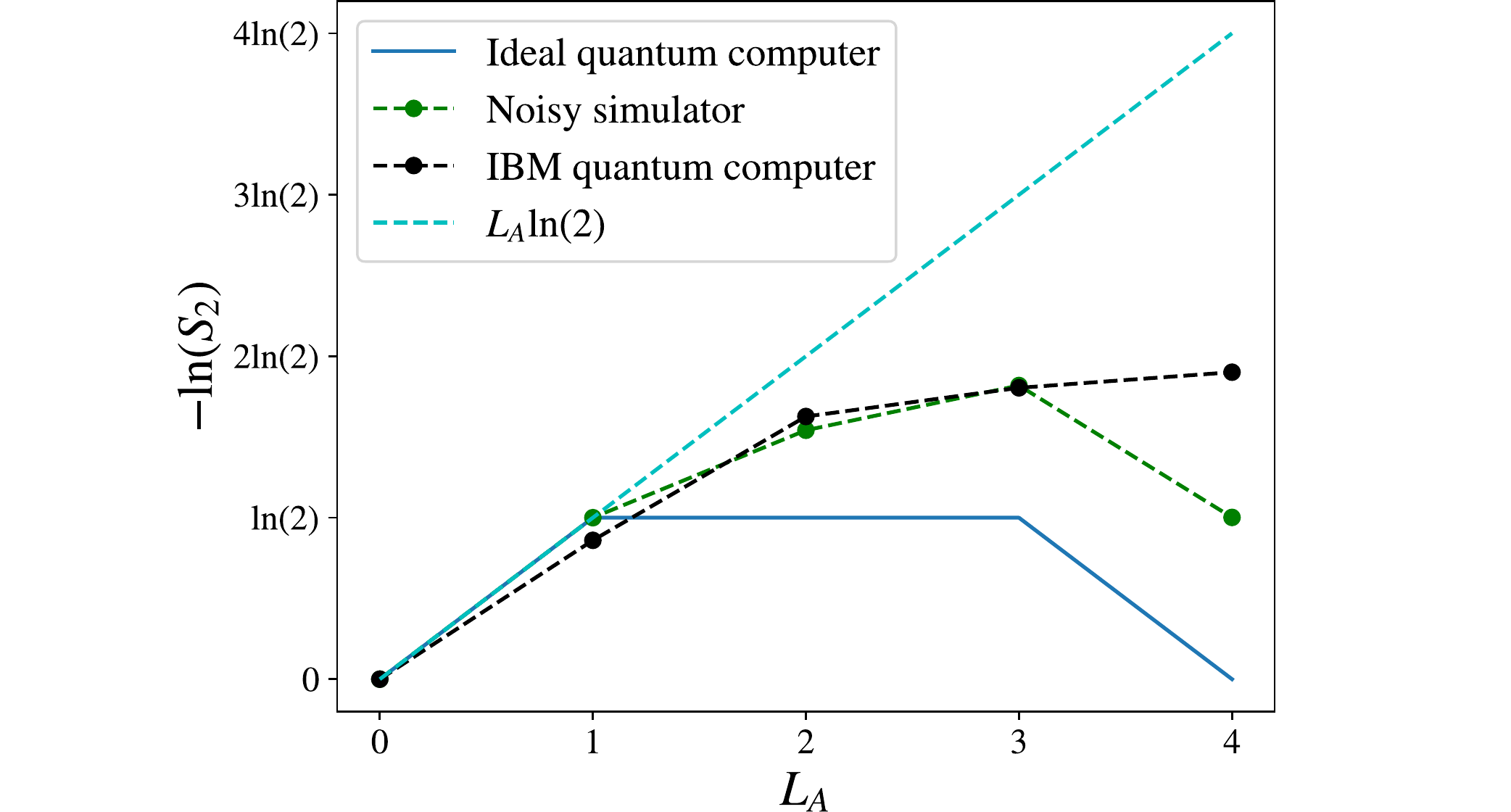}
	\caption{Realization and characterization of a 
	cluster
		state $\ket{\psi_{\rm cluster}}$ with $L=4$ qubits. Second 
		\Renyi entropy
		as a function of the subsystem size $L_A$.}
	\label{fig:entropy}	
\end{figure}

\mysection{Symmetry-resolved entropies}
As mentioned in the introduction, we use symmetry-resolved reduced 
density matrices, $\tilde\rho_A$, to identify the SPT nature of the
cluster state. A direct measure of these matrices (state tomography) 
requires an exponentially large number of measurements. We overcome 
this difficulty by addressing the moments of these matrices, 
$\tilde{S}_n = {\rm Tr}[\tilde{\rho}^n_A]$, which can be measured 
by realizing $n$ copies of the state \cite{daley2012measuring,%
pichler2013thermal,islam2015measuring,%
johri2017entanglement,cornfeld2018imbalance,cornfeld2019measuring,%
bonsignori2019symmetry,feldman2019dynamics,murciano2019symmetry,%
tan2019particle}. Specifically, for $n=2$, this approach is based 
on the identity
\begin{align} {\rm Tr}[\rho^2] = {\rm Tr}[\rho\otimes\rho~ {\rm SWAP}].
\label{eq:SWAP}
\end{align}
Here, $\rho\otimes\rho$ is the combined state of two independently 
prepared copies of a state, and the operator SWAP swaps the states 
of the two copies. By applying the SWAP operator only to the subsystem $A$, 
one can compute the purity of $A$, ${\rm Tr}[\rho_A^2]$. Finally, if the 
SWAP operator is measured along with the projector to the conserved sectors, 
one can directly obtain the symmetry-resolved entropy
$\tilde{S}_n$~\cite{cornfeld2018imbalance, xavier2018equipartition,
	barghathi2019operationally} 
\footnote{Mathematically, a proof of the equivalence between the symmetry 
sectors requires to access the full ES, demanding a number of measurements 
that grows with the size of the Hilbert space. However, in practice, it is 
often sufficient to check that the symmetry exists in the first few moments 
to demonstrate the SPT nature of the state.}.

\begin{figure*}[t]
	\includegraphics[width=1\textwidth]{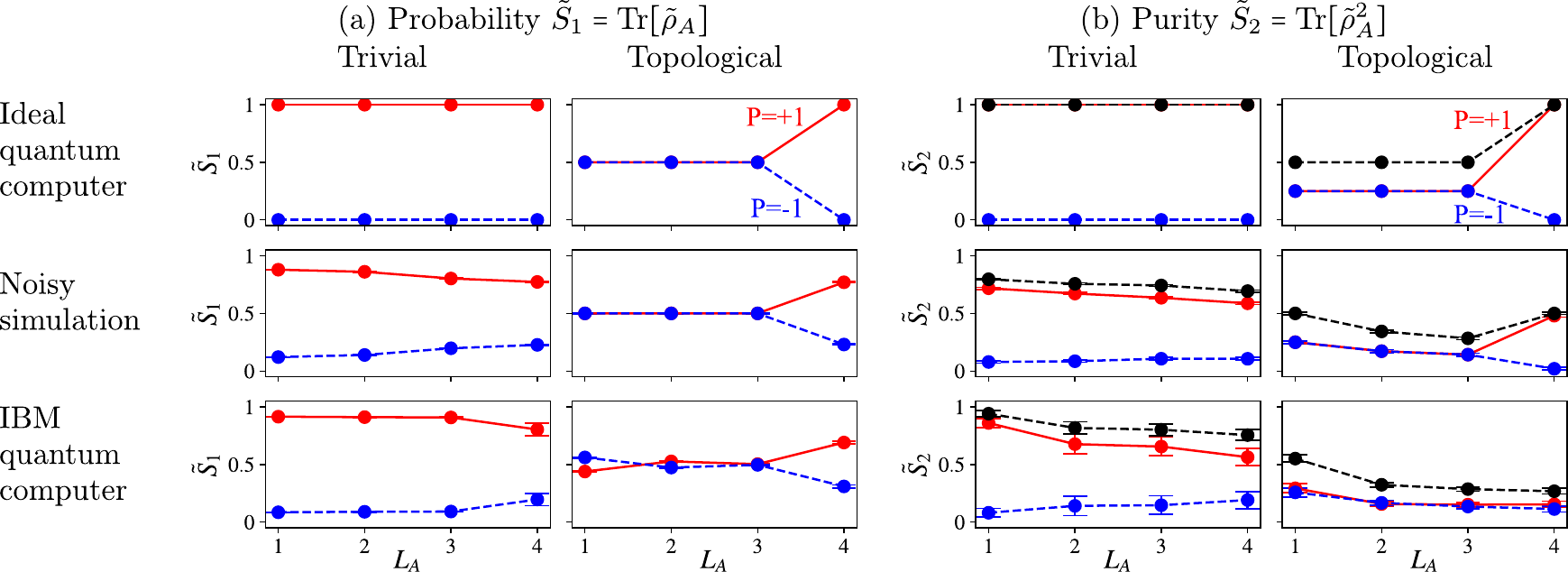}
	\caption{Symmetry-resolved entanglement measures
		$\tilde{S}_1$ and $\tilde{S}_2$, for the trivial state 
		$|\psi_{\rm
			trivial}\rangle$ and for the topological state 
			$\ket{\psi_{\rm cluster}}$.
		The $P=+1$, $P=-1$, and total 
		contributions are
		shown in red, blue and black, respectively.}
	\label{fig:resolved}
\end{figure*}

To implement these ideas on a quantum computer, we create two copies 
of the cluster state with $L=4$ qubits, using twice the circuit of
Fig.~\ref{fig:circuits}(a). Next, we measure the SWAP operator on each 
pair of qubits of the two copies, using the quantum circuit introduced by
Refs.~\cite{garcia2013swap,cincio2018learning}, see Fig.~\ref{fig:circuits}(b)
\footnote{All circuits used in this can be found in the 
Supplemental Materials}. 
By repeatedly measuring the output of the circuit, we infer the
expectation values of the products of the SWAP operators of each site 
of a subsystem A, which correspond to $S_2={\rm Tr}[\rho_A^2]$. In 
Fig.~\ref{fig:entropy} we plot $-{\rm ln} S_2$, also known as the 
{\it second \Renyi entropy}, as a function of the subsystem size $L_A$. 
The result of this calculation matches the known properties of the cluster
state with open boundary conditions: For any $0<L_A<L$, $\rho_A$ has 2
identical eigenvalues $\lambda=1/2$, one for each sector of the 
symmetry $P$, and one has $S_{2} = 1/2$. Importantly, for $L_A=L$ one has 
 $S_2={\rm Tr}[\rho^2]=1$, indicating 
that the system is pure.
	
We now turn to symmetry-resolved measurements, which unveil the SPT 
nature of a state. The first moment, $\tilde{S}_1 = {\rm Tr}[\tilde\rho_A]$, is 
simply the probability to find a subsystem in a specific sector of the 
symmetry. To compute the second moment, we develop a method to measure the 
product of the SWAP and $P$ operators, where $P$ acts on one copy only, see 
Fig.~\ref{fig:circuits}(c). This method can be generalized to richer symmetries with a larger number of topological phases, see section A of the Supplemental Materials for details. The results of these 
calculations are shown in the upper panel of Fig.~\ref{fig:resolved}: 
For the trivial state, the entire weight lies in the even parity  sector, 
$P=+1$. For the cluster state, the full system ($L_A=L$) is still an 
eigenvector of $P$ with $P=+1$. In contrast, smaller subsystems ($L_A<L$) 
occupy with equal probabilities the sectors $P=+1$ and $P=-1$, in 
agreement with the topologically-protected degeneracy of the 
symmetry-resolved reduced density matrices.

\mysection{Noisy SPT states} To understand actual experiments, it is necessary to study the effect of noise 
on topological states. Several earlier works addressed this question by
extending the topological classification of pure states to density matrices 
\cite{diehl2011topology,viyuela2014uhlmann,van2014classification,%
	andersson2016geometric,linzner2016reservoir,gong2018topological,%
	bardyn2018probing,goldstein2018dissipation,asorey2019topological}. 
Here, we focus on the effect of noise on the degeneracies of the 
ES, as probed by symmetry-resolved reduced density matrices. We 
define a noise source to be symmetry preserving if it preserves this 
degeneracy (and vice versa). See Refs.~\cite{buca2012note,baumgartner2008analysis,albert2014symmetries} and Sec. B of the Supplemental Materials for a formal definition. 

Let us consider the results of a noisy simulation, obtained using
QISKIT AER \cite{Qiskit}. The simulator computes the
evolution of the density matrix by taking into account realistic noise
sources in terms of Kraus operators. The parameters used in the simulation
are determined by direct measurements of the success probability of the
gates in the physical system \footnote{We used the noise model that 
matches the experiments' date.}. Interestingly, all noise sources present 
in this simulation are symmetry preserving \cite{QISKITnoise}, with the 
exception of a measurement bias that leads to a systematic error towards 0 
outcomes. To study the effects of symmetry preserving noise sources, we 
manually eliminate this bias from the simulations. In this case, if the 
system is prepared in an SPT state belonging to the same universality class 
as the cluster state, the noise does not lift the ES degeneracies.

We first consider the effects of noise on $S_2={\rm Tr}[\rho^2]$, see
Fig.~\ref{fig:entropy}. In the presence of noise, the state is not 
pure and the second \Renyi entropy of the full system is $\approx 
{\rm ln(2)}$. This value is significantly smaller than the maximally 
allowed value of $4{\rm ln(2)}$, indicating that the output of the 
simulation is not trivial. The slope of the entropy changes in the 
second half of the chain, as in the ideal quantum computer. To study the SPT
properties of this noisy state, we compute symmetry-resolved 
quantities, see Fig. \ref{fig:resolved}. For the trivial state, we find that 
both the probability and the symmetry-resolved purity are larger for $P=+1$ 
than for $P=-1$. In contrast, in the cluster state the probabilities and 
purities are identical for the two sectors for all $L_A<L$. Remarkably, the 
total system ($L_A=L$) is mostly in the $P=+1$ state, confirming that 
the system is targeting the correct pure state.

Using the same QISKIT package \cite{Qiskit}, we performed the
same calculations on the 15-qubit Melbourne IBM quantum
computer, using 150 runs with 8192 measurements each
\footnote{The experiments were performed on December 26-28, 2019.}.
This computer has 15 qubits organized in a ladder structure, with physical 
two-qubit gates between nearest neighbors only. This structure is ideal for 
the circuit under the present consideration: we realize the two copies of 
the cluster states on the two parallel chains that form the ladder, and use 
the rungs to realize the SWAP operators. The results obtained in the actual 
computer are similar to those observed in the simulator: although the purity 
of the cluster state is not ideal, our symmetry resolved probes still 
correctly identify its SPT nature. One interesting difference between the 
quantum computer and the noisy simulator can be observed in the symmetry 
resolved probes of small subsystems, $L_A=1,2$. In the actual computer, the two 
sectors show small, but statistically significant, differences. We 
identify these errors as due to symmetry-breaking noise sources, such as the
aforementioned measurement bias, which were absent in the simulation but 
present in the physical system. This bias also explains why the \Renyi
entropy of the $L_A=1$ subsystem (Fig.~\ref{fig:entropy}) is smaller
than 1/2, see section C of the Supplemental Materials. Our results demonstrate that 
topological arguments can be used to characterize the main sources of 
errors and classify them according to their symmetry.

\begin{figure}[t]
	\centering
	\vspace{-0.5cm}
	\includegraphics[width=0.48\textwidth]{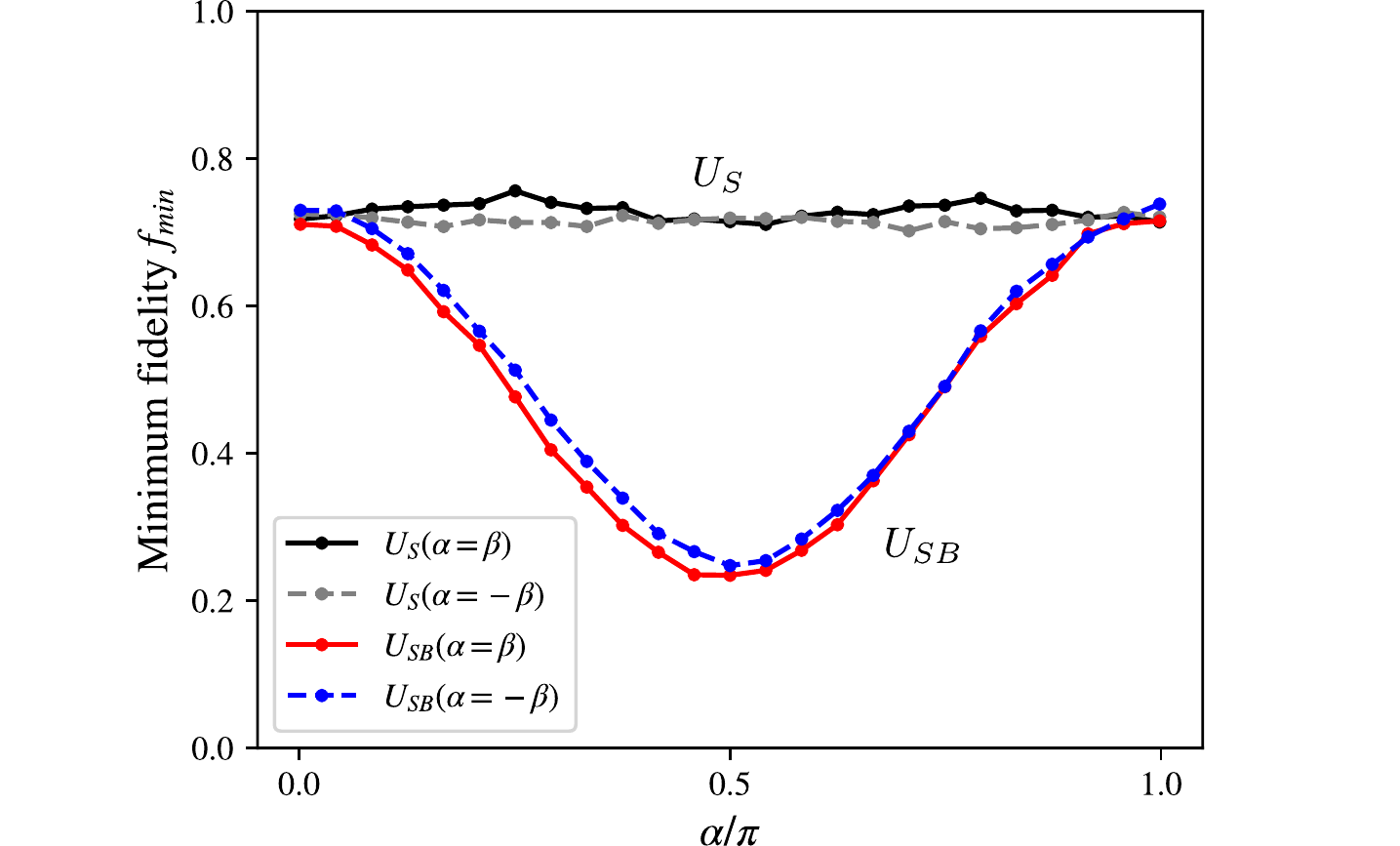}
	
	\caption{Fidelity of the MBQC teleportation
		algorithm under the influence of symmetry-(non)preserving 
		perturbations.
		Each data point represents the minimal fidelity with respect 
		to 6 initial
		states (see section E of the Supplemental Materials for the raw data).}	
		\label{fig:refael}	
\end{figure}

\mysection{Measurement-based wire protocol} We now turn to the 
experimental
realization of the symmetry-protected wire protocol
\cite{else2012symmetry}. In this protocol, a general quantum state 
is encoded in one boundary of the spin chain. The state is, then, 
shuttled
to the other boundary in a teleportation-like fashion, by local
measurements of the spins along the chain. We apply this protocol to a
family of SPT states with $\mathbb{Z}_2\times \mathbb{Z}_2$ symmetry, 
which
contains the 1D cluster state as a special case. All states in the 
family
possess the same SPT order and, hence, have the same capacity to 
transmit
one-qubit-worth of quantum information. Our goal is to verify the
robustness of the protocol against variation within the phase.

For our implementation on an IBM quantum computer we use the $L=4$ 
cluster
state $\ket{\psi_{\rm cluster}}$ described above. The corresponding
$\mathbb{Z}_2 \times \mathbb{Z}_2$ symmetry is generated by
$P_{\text{odd}} =\prod_{i=1,3}h_i=X_1X_3Z_4$
and
$P_{\text{even}} =\prod_{i=2,4}h_i=Z_1X_2X_4$, where $h_i$ are defined 
in 
Eq.~\ref{eq:Hcluster}.
The family of SPT states is created applying either 
symmetry-preserving 
unitaries 
$U_S(\alpha,\beta) = e^{i\beta Z_1 X_2 Z_3}e^{i \alpha X_3}$, 
or symmetry-breaking unitaries 
$U_{SB}(\alpha,\beta) = e^{i\beta Z_1 X_2 Z_3}e^{i \alpha Y_3}$ to 
$\ket{\psi_{\rm cluster}}$.
In the former case all resource states respect the 
$\mathbb{Z}_2 \times \mathbb{Z}_2$ symmetry and can be continuously 
connected in a symmetry-respecting
fashion to the cluster state. In the latter case, the symmetry is 
broken
and 
computational uniformity is not guaranteed.

Next, we introduce another qubit realizing the input state 
$\ket{\psi_{in}}$ and teleport 
it into the wire by performing a measurement in the 2-qubit cluster 
basis
(a 
locally rotated Bell basis, $\{\ket{+0}\pm\ket{-1}\}$ ) on
$\ket{\psi_{in}}$ 
and the first qubit of the spin chain, see Fig.~\ref{fig:circuits}(d). 
This particular measurement is chosen to be compatible with the 
MBQC wire protocol,
consisting of local measurements in the X-basis of the remaining 
qubits,
and classically controlled Pauli correction 
depending on the measurement
outcomes. 
Fig. \ref{fig:refael} shows the experimentally measured minimum 
fidelity 
$f_{min} = \min_i \,\bra{\psi_{in}^i} 
\rho_{out}^{exp}\ket{\psi_{in}^i}$
for 
six different input states $\ket{\psi_{in}^i}$ and the Pauli-corrected
output 
state $\rho_{out}^{exp}$ resulting from the wire protocol, for the 
choices 
$\beta=\pm \alpha$
in both the symmetric and the symmetry-breaking case, see also section 
E of
the 
Supplemental Materials. We find that the transmission
fidelity is constant as a function of $\alpha$ in the 
symmetry-respecting
case. 
In the symmetry-breaking case, the transmission fidelity is 
non-constant 
as the resource state is varied.

\mysection{Conclusion} In this paper we devised and implemented
experimentally two methods to identify the SPT nature of the cluster 
state
on a quantum computer. The first algorithm stems from the observation 
that
in SPT states, the reduced density matrix $\rho_A$ is formed by 
identical
blocks that correspond to different sectors of the underlying 
symmetry. The
flexibility of the quantum computer makes it possible to directly 
probe the
moments of density matrices by projecting the quantum state into the
different symmetry sectors. The realization of this algorithm on both a
quantum simulator and on a IBM quantum computer allowed us to study the
impact of time dependent noise on the SPT order of the state. In
particular, we found that while most of realistic noise sources are
symmetry preserving, the systematic measurement bias of the physical
machine breaks this symmetry. Its effects are, however, small enough to
enable us to identify the SPT nature  of the cluster state. 

The second way
to characterize the SPT order of the cluster states consists of using 
them
as a buffer for MBQC teleportation. The fidelity 
of
this protocol is unaffected by symmetry preserving terms, and vice 
versa
for symmetry breaking terms, allowing us to identify the SPT order of 
the
cluster states. The equivalence between these two methods demonstrate a deep relation between two separate fields of science---condensed matter physics and quantum information theory.

Our work has important implications for the modeling of noisy
intermediate-scale quantum computers. We demonstrated that topological
arguments are an efficient tool to identify and classify noise sources 
in
quantum computers. This information can be used to improve the 
performance
of quantum computers, for example, by gauging the measurement 
apparatus to
take into account systematic errors. From a fundamental perspective, we
identified sufficient conditions under which a noisy quantum state can
retain its SPT properties. This aspect may have implications for 
quantum
computations: for pure states, it was shown that the  classification 
of SPT
phases is in one-to-one correspondence with the possibility to use it 
as a 
resource for one-way-quantum computer. Although this question deserves
further investigation, we conjecture that this link extends to noisy
systems as well. 

{\bf Acknowledgments} We acknowledge useful discussions with Yael
Ben-Haim, Moshe Goldstein, Joe Jackson, Yuval Tamir, Ari Turner, Hannes
Pichler, Frank Pollmann. We acknowledge support from ARO 
(W911NF-20-1-0013)
(RR and ES). This work is supported by the Israel Science Foundation,
grants number 151/19 (DA and EGDT), 154/19 (DA, YN, ES and EGDT). We acknowledge the use of IBM Quantum services for this work. The views expressed are those of the authors, and do not reflect the official policy or position of IBM or the IBM Quantum team.

\section*{Supplemental Materials}

\subsection{Quantum algorithm to compute the symmetry resolved purity}
\label{sec:purity_algorithm}
The symmetry resolved purity of the subsystem $A$ of size $L_A<L$ is defined by $\tilde{S}_2(P) = {\rm Tr}[\rho_A^2 \Pi_A(P)]$, where $\Pi_A(\pm 1)=(1\pm Y_1X_2...X_{L_A})/2$ is the projection over the $P=\pm1$. We implement this circuit by taking the average between the expectation values of ${\rm{Tr}} [\rho_A^2]$ and ${\rm{Tr}} [\rho_A^2 Y_1X_2...X_{L_A}]$. To compute the latter, we implement two copies of the same state, according to Eq.~(1). For simplicity, let us focus on a single qubit $i$, where the operator ${\rm{Tr}} [\rho_A^2 X_i]$ can be written as ${\rm{Tr}} [ \rho_2 (X_i \otimes I) ~{\rm{SWAP_i}} ]$ and ${\rm SWAP}_i$ swaps the two copies of the qubit $i$. The operator $O_i=(X_i \otimes I_i) ~{\rm{SWAP_i}}$ is unitary, $O_i^\dagger O_i=1$ with eigenvectors $\{ |++ \rangle ,|-- \rangle , \frac{|+- \rangle + i |-+ \rangle}{\sqrt{2}},  \frac{|+- \rangle - i |-+ \rangle}{\sqrt{2}}$ and eigenvalues $\{ \lambda_i\} =\{1,-1,i ,-i \}$. This local basis change is performed in Fig.~1(c) using the Z basis and needs to be rotated to the X basis for $i>2$ (or the Y basis for $i=1$). To obtain ${\rm{Tr}} [\rho_A^2 \Pi_A]$, after performing a measurement on each pair of copies and classically recording the appropriate eigenvalue $\lambda_i$, we perform a quantum average over $\prod_{i=1}^{L_A} \lambda_i$.

This method generalizes for any moment $n$ and for general finite Abelian symmetry (such as $Z_N$), hence generalizing the symmetry-resolved entanglement protocols of Refs.~\cite{cornfeld2018imbalance,cornfeld2019entanglement} to qubits. The general projections for finite Abelian group $G$ are $P_k =\frac{1}{|G|} \sum_{g_i \in G} \chi_k(g_i) U(g_i)$ where $\chi_k(g_i)$ are known as the multiplicative characters (homomorphisms from the group $G$ to $\mathbb{C}$) of the group $G$ and can be found in the group theory literature and $U(g_i)$ are the local symmetries $U(g_i) = \otimes U_{i}$ decomposed on their respective sites. The moments of the $k$ sector are given by $\mathrm{Tr}[\rho^n P_k]$, and so we only need to calculate $\mathrm{Tr}[\rho^n U(g_i)]$ and sum it accordingly. This can be done by going to the $n$ copy representation (a.k.a the SWAP test) and from the locality of $U(g_i)$ we only need to apply a transformation (local basis change from the $U_i$ basis to the computational basis accompanied with basis change to the eigenbasis of $C_n (I\otimes I \otimes \dots \otimes Z)$, where $C_n$ is the cyclic permutation of size $n$) on groups of $n$ qubits, where each of these qubits is taken from another copy. The gate complexity does not exceed $O(f(n)L)$, where $L$ is the number of qubits in one copy and $f(n)$ is the complexity of the transformation to the eigenbasis of $C_n (I\otimes I \otimes \dots \otimes Z)$. The number of circuits scales with $|G|$, the number of symmetry sectors in $G$. Optimizations can further reduce the gates (such as combining the SWAP with the local basis as was done here). Using this general method, one can calculate the symmetry-resolved moments of the entanglement for any finite-size Abelian symmetry $G$.

\subsection{Formal definition of symmetry preserving noise sources}
\label{sec:noise_definition}
A formal definition of symmetry preserving noise sources can be given by introducing an operator $T_A$, which acts on a subsystem $A$ and maps the different sectors of the symmetry among themselves. In a SPT state, all symmetry-resolved reduced density matrices are identical and hence $[T_A,\rho_A]=0$. In the example of the cluster state the operators $T_A$ flip the edge spins $X_{1}$ and $X_{L_A}$ and are given by $Z_{1}$ and $Z_{L_A}$. A generic noise map $\Phi:\rho_A\to\rho_A'$ is then said to be symmetry-preserving if it preserves the property $[T_A,\rho'_A]=0$. Specifically, we focus on noise sources that can be described by the Kraus operators according to
\begin{equation}
\Phi:\rho \to \rho'=\sum_i K_i \rho K_i^\dagger
\end{equation}
with the normalization condition $\sum_i K_i^\dagger K_i = I$, where $I$ is the identity matrix. A trivial example of a symmetry-preserving noise is dephasing, described by the Kraus operators $K_1=\sqrt{1-p}I$ and $K_2=\sqrt{p}Z_i$. Both operators conserve $Z_i$ and commute with $T_A$. A non-trivial example is given by the depolarizing noise with $K_1=[(1+\sqrt{1-p})I-(1-\sqrt{1-p})Z_i]/2$ and $K_2=\sqrt{p}\sigma^-_i$. These operators do not conserve $Z$ and, hence, do not commute with $T_A$. However, because $\sigma^-_i$ commutes with the product of two $K_i$, if $[\rho,Z_i]=0$ then $[\rho',Z_i]=0$ leading to symmetry preservation. These examples highlight the difference between conserved quantities and symmetries:  a conserved quantity is always a symmetry, but not vice versa (see Refs.~\cite{buca2012note,baumgartner2008analysis,albert2014symmetries} for an introduction).

\vspace{1em}

\subsection{A simple model of the measurement bias}
\label{sec:noise_model}
A natural candidate for the symmetry-breaking noise observed in the quantum computer is a systematic error present in the measurement device, giving preference to state 0 with respect to state , or vice versa. The existence of this error explains why the second \Renyi entropy $-{\rm ln}[S_2]$ at $L_A=1$ is smaller than ${\rm ln2}$, see Fig. 3 in the main text: If we assume that the output qubits are random variables with probabilities $0.5+\epsilon$, $ 0.5-\epsilon$ for 0, 1 respectively, we obtain ${-\rm ln}S_2 = - L_A {\rm ln}\left[(0.5+\epsilon)^2+2(0.5+\epsilon)(0.5-\epsilon)-(0.5-\epsilon)^2\right] \approx L_A ( {\rm ln}2-4\epsilon)$. The proof lies in the fact that usual SWAP test maps a singlet on a pair of qubits from both copies into a pair of ones in both copies, resulting in the above expression as every such measurement of singlet will multiply the overall SWAP eigenvalue by $-1$. In the same model, the difference between the even and odd probabilities decreases exponentially as $|\tilde{S}_1(P=+1)-\tilde{S}_1(P=-1)| = |(0.5+\epsilon) - (0.5-\epsilon)|^{L_A} = |2\epsilon|^{L_A}$. These expressions are in qualitative agreement with the experimental observations for  $\epsilon \approx 7\%$, see Figs. 3 and 4 in the main text.

\subsection{Reduced density matrix of the cluster state}
The properties of the cluster states can be derived by noting that all the stabilizers $h_i\equiv Z_{i-1}X_iZ_{i+1}$, commute among each other. Because $h_i$ is Hermitian and squares to 1, its eigenvalues are $\pm 1$. For $L$ qubits with periodic boundary conditions (and even $L$, in consistency with the $Z_2\times Z_2$ symmetry), the $2^L$ common eigenvectors of the $h_i$'s form an orthonormal basis. The ground state of the Hamiltonian $H_{\rm cluster}$ corresponds to the state satisfying $h_i|\psi_{\rm cluster}\rangle=|\psi_{\rm cluster}\rangle$ for all $i$. 

We now use this construction to derive the reduced density matrix $\rho_A = {\rm Tr}_B[|\psi_{\rm cluster}\rangle\langle\psi_{\rm cluster}|]$. Specifically, we consider as the subsystem $A$ the qubits $i$ with $1\leq i\leq L_A$.  The reduced density matrix is obtained from the Schmidt decomposition  $|\psi_{{\rm{cluster}}}  \rangle =\sum_{i} \lambda_i |\psi^A_{i}  \rangle |\psi^B_{i}  \rangle$ as $\rho_A = \sum_i |\lambda_i|^2 |\psi^A_{i}  \rangle \langle \psi^A_{i}| $. For a SPT phase it is convenient to perform the Schmidt decomposition in terms of edge states~\cite{pollmann2010entanglement,fidkowski2010entanglement}. Here, the left edge $(\ell)$ state of region $A$ consists of Pauli operators $\mathcal{Z}_\ell=Z_1$ and $\mathcal{X}_\ell=X_1 Z_2$ and for the right edge $\mathcal{Z}_r=Z_{L_A}$ and $\mathcal{X}_r=Z_{L_A-1}X_{L_A}$. It is sufficient to perform the Schmidt decomposition on the subspace spanned by the few (four) stabilizers $h_{i}$ that connect $A$ and $B$ across the two entanglement cuts. One finds that near each entanglement cut the joint stabilizer eigenstates are Bell states of the edge spins across each entanglement cut. Thus, in the basis $| \alpha = \pm 1 \rangle$ ($| \beta = \pm 1 \rangle$) of eigenstates of edge spins $\mathcal{X}_\ell$ ($\mathcal{X}_r$), we have $\rho_{A} =\frac{1}{4} \sum_{\alpha , \beta} |\alpha , \beta \rangle \langle \alpha , \beta |$. This expression indicates that $\rho_A$ has 4 identical eigenvalues, $\lambda_i=1/4$. Crucially, these edge operators represent the symmetry within the ground state. Using the fact that all stabilizers within the bulk of the subsystem satisfy $h_i=1$, we have (for $L_A$ odd and an obvious modification for  $L_A$ even) $P_{\rm{odd}} =\mathcal{X}_\ell \mathcal{X}_r= \alpha \beta$ and $P_{\rm{even}} = \mathcal{Z}_\ell \mathcal{Z}_r$. Diagonalizing the symmetries, we see that each eigenvalue belongs to a different sector of the symmetries $(P_{\rm even},P_{\rm odd})=(\pm1,\pm1)$. As expected for a SPT state, one obtains equal contributions from all symmetry sectors. For open-boundary conditions, there is only one edge in the Schmidt decomposition, resulting in $\rho_{A} =\frac{1}{2} \sum_{\beta} |\beta \rangle \langle \beta |$, 
which has 2 identical eigenvalues $\lambda_i=\frac{1}{2}$.	Finally, for the topologically trivial state  $|\psi_{\rm trivial}\rangle$, which is a product state, the reduced density matrix has a single eigenvalue $\lambda=1$ belonging to the sector $(P_{\rm even},P_{\rm odd})=(1,1)$.

\subsection{Fidelity of the quantum teleportation algorithm}
\label{sec:refael}

\begin{figure*}[h]
	\centering
	\includegraphics[width=0.8\textwidth]{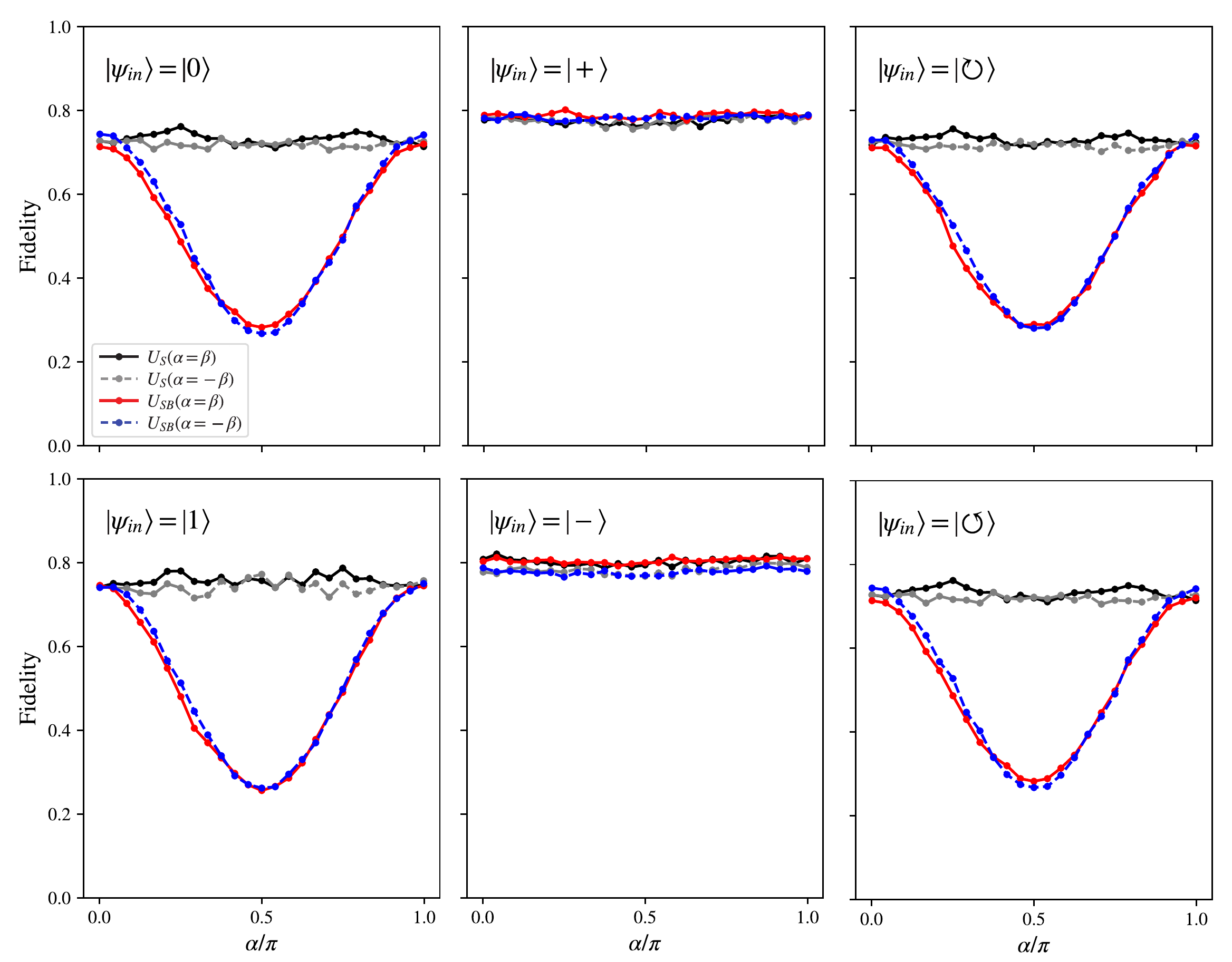}
	\caption{Fidelity of the teleportation algorithm for six different initial states.}
	\label{fig:comb}
\end{figure*}

The fidelity the quantum teleportation algorithm is defined as $F = |\langle \psi_{in}|\psi_{\rm out}'\rangle|^2$, where $\ket{\psi_{\rm out}'} = U\ket{\psi_{\rm out}}$, $U=Z_{\rm out}^{q_1}X_{\rm out}^{q_2}Z_{\rm out}^{q_3}X_{\rm out}^{q_4}$, and $\{q_1,~q_2,~q_3,~q_4\}$ are the measured values of the wire qubits in the $x$ basis \cite{else2012symmetry}. For a mixed output state characterize by the density matrix $\rho_{\rm out}$, the fidelity generalizes to $F = \bra{\psi_{\rm in}} U\rho_{\rm out}U^\dagger \ket{\psi_{\rm in}}$. We measure $\rho_{\rm out}$ by full state tomography, i.e. by measuring the expectation values of $X,~Y$ and $Z$ and using $\rho_{\rm out}=\left(1+\langle X_{\rm out}\rangle \sigma^x + \langle Y_{\rm out}\rangle \sigma^y + \langle Z_{\rm out}\rangle \sigma^z\right)/2$. The fidelity of the quantum teleportation algorithm is computed for 6 different initial stats: 
\begin{align}
\ket{0}&\,, \quad
\ket{1}\,, \quad
\ket{+}\,, \quad
\ket{-}\,, \quad\nonumber\\
\ket{\lcirclearrowright}& = \frac{1}{\sqrt{2}}(\ket{0}+i\ket{1})\,, \quad
\ket{\rcirclearrowleft} = \frac{1}{\sqrt{2}}(\ket{0}-i\ket{1}).
\end{align} The results of our algorithm for each individual initial state are shown in Fig.~\ref{fig:comb}. Each data point is obtained by averaging over 8192 measurements.

\clearpage
\begin{widetext}
	\subsection{Quantum circuits used in this article}
	\label{sec:codes}

	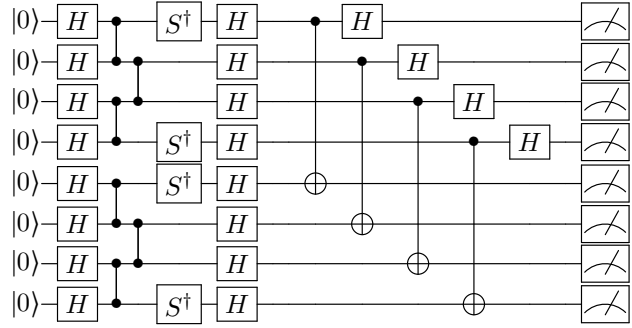
\begin{figure}[h]
		\centering
		\begin{minipage}[t]{1\textwidth}
			\Qcircuit @C=0.6em @R=0.1em 
			{ 
				\\ \ket{0}&& \gate{H} & \ctrl{1}  & \qw       &\gate{S^\dagger}&\gate{H} &\qw &\qw  &\ctrl{4} & \gate{H} & \qw & \qw& \qw & \qw& \meter
				\\ \ket{0}&& \gate{H} & \ctrl{-1} & \ctrl{1}  &\qw             &\gate{H} &\qw &\qw  &\qw &\ctrl{4} & \gate{H} & \qw & \qw & \qw& \meter
				\\ \ket{0}&& \gate{H} & \ctrl{1}  & \ctrl{-1} &\qw             &\gate{H} &\qw &\qw  &\qw &\qw  &\ctrl{4} & \gate{H} & \qw & \qw& \meter
				\\ \ket{0}&& \gate{H} & \ctrl{-1} & \qw       &\gate{S^\dagger}&\gate{H} &\qw &\qw  &\qw &\qw  &\qw &\ctrl{4} & \gate{H} & \qw & \meter
				\\ \ket{0}&& \gate{H} & \ctrl{1}  & \qw       &\gate{S^\dagger}&\gate{H} &\qw &\qw  &\targ & \qw & \qw & \qw & \qw& \qw& \meter
				\\ \ket{0}&& \gate{H} & \ctrl{-1} & \ctrl{1}  &\qw             &\gate{H} &\qw &\qw  &\qw &\targ & \qw & \qw & \qw & \qw& \meter
				\\ \ket{0}&& \gate{H} & \ctrl{1}  & \ctrl{-1} &\qw             &\gate{H} &\qw &\qw  &\qw &\qw  &\targ & \qw & \qw & \qw& \meter
				\\ \ket{0}&& \gate{H} & \ctrl{-1} & \qw       &\gate{S^\dagger}&\gate{H} &\qw &\qw  &\qw &\qw  &\qw &\targ & \qw & \qw & \meter
			}
		\end{minipage}
		\caption{Circuit for measuring the purity $S_A$.}
	\end{figure}

	\begin{figure}[h]
		\centering
		\begin{minipage}[t]{1\textwidth}
			\Qcircuit @C=0.6em @R=.1em 
			{ 
				\\ \ket{0}&& \gate{H} & \ctrl{1}  & \qw       &\gate{S^\dagger}&\gate{H} &\meter
				\\ \ket{0}&& \gate{H} & \ctrl{-1} & \ctrl{1}  &\qw             &\gate{H} &\meter
				\\ \ket{0}&& \gate{H} & \ctrl{1}  & \ctrl{-1} &\qw             &\gate{H} &\meter
				\\ \ket{0}&& \gate{H} & \ctrl{-1} & \qw       &\gate{S^\dagger}&\gate{H} &\meter
			}
		\end{minipage}
		\caption{Circuit for measuring the symmetry-resolved probabilities $\tilde{S}^A_1$.}
	\end{figure}
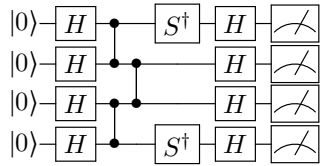

	\begin{figure*}[h]
		\centering
		\begin{minipage}[t]{1\textwidth}
			\Qcircuit @C=0.6em @R=.1em 
			{ 
				\\ \ket{0}&& \gate{H} & \ctrl{1}  & \qw       &\qw &\gate{S^\dagger}& \gate{H} & \qw  &\targ    &\qw & \ctrl{4} & \targ &\qw &\qw &\qw &\qw &\qw &\qw &\qw &\qw &\qw &\qw &\qw &\qw & \meter
				\\ \ket{0}&& \gate{H} & \ctrl{-1} & \ctrl{1}  &\qw &\qw& \gate{H} & \qw  &\qw &\qw &\qw &\qw &\targ    &\qw & \ctrl{4} & \targ &\qw &\qw &\qw &\qw &\qw &\qw &\qw &\qw & \meter
				\\ \ket{0}&& \gate{H} & \ctrl{1}  & \ctrl{-1} &\qw &\qw& \gate{H} & \qw  &\qw &\qw &\qw &\qw &\qw &\qw &\qw &\qw &\targ    &\qw & \ctrl{4} & \targ &\qw &\qw &\qw &\qw & \meter
				\\ \ket{0}&& \gate{H} & \ctrl{-1} & \qw       &\qw &\gate{S^\dagger}& \gate{H} & \qw  &\qw &\qw &\qw &\qw &\qw &\qw &\qw &\qw &\qw &\qw &\qw &\qw &\targ &\qw & \ctrl{4}  &\targ    & \meter
				\\ \ket{0}&& \gate{H} & \ctrl{1}  & \qw       &\qw &\gate{S^\dagger}& \gate{H} & \qw  &\ctrl{-4}& \gate{S^\dagger} & \gate{H} & \ctrl{-4}&\qw &\qw &\qw &\qw &\qw &\qw &\qw &\qw &\qw &\qw &\qw &\qw &   \meter
				\\ \ket{0}&& \gate{H} & \ctrl{-1} & \ctrl{1}  &\qw &\qw& \gate{H} & \qw  &\qw &\qw &\qw &\qw &\ctrl{-4}& \gate{S^\dagger} & \gate{H} & \ctrl{-4}&\qw &\qw &\qw &\qw &\qw &\qw  &\qw &\qw &\meter
				\\ \ket{0}&& \gate{H} & \ctrl{1}  & \ctrl{-1} &\qw &\qw& \gate{H} & \qw  &\qw &\qw &\qw &\qw &\qw &\qw &\qw &\qw &\ctrl{-4}& \gate{S^\dagger} & \gate{H} & \ctrl{-4}&\qw &\qw &\qw &\qw &   \meter
				\\ \ket{0}&& \gate{H} & \ctrl{-1} & \qw       &\qw &\gate{S^\dagger}& \gate{H} & \qw  &\qw &\qw &\qw &\qw &\qw &\qw &\qw &\qw &\qw &\qw &\qw &\qw &\ctrl{-4}& \gate{S^\dagger} & \gate{H} & \ctrl{-4} &   \meter
			}
		\end{minipage}
		\caption{Circuit for measuring the symmetry-resolved purities $\tilde{S}^A_2$.}
	\end{figure*}
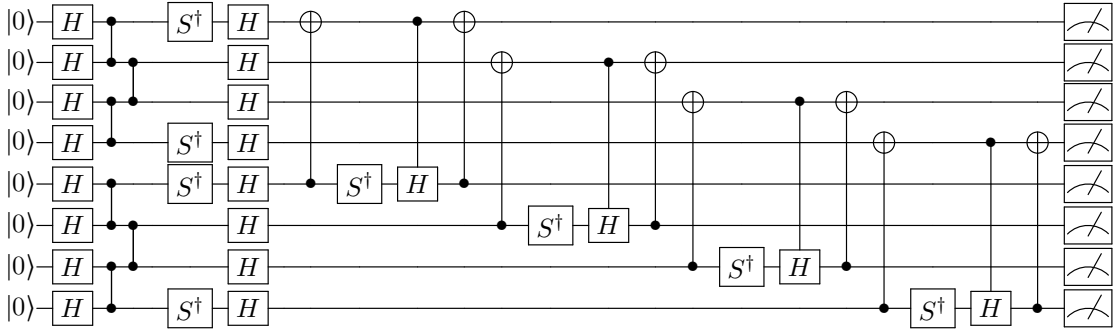
	
	\begin{figure}[h]
		\centering
		\begin{minipage}[t]{1\textwidth}
			\Qcircuit @C=0.6em @R=.1em {
				&\lstick{\ket{\psi_{in}}}&\qw      &\ctrl{1} &\qw      &\qw                  &\qw       &\qw        &\qw
				&\qw                     &\qw      &\gate{H}&\meter
				\\
				&\lstick{\ket{0}}        &\gate{H} &\ctrl{-1}&\ctrl{1} &\qw                  &\ctrl{1}  &\qw        &\qw
				&\qw                     &\ctrl{1} &\gate{H}&\meter
				\\
				&\lstick{\ket{0}}        &\gate{H} &\ctrl{1} &\ctrl{-1}&\qw                  &\ctrl{-1} &\ctrl{1}   &\gate{e^{-i\beta X}}
				&\ctrl{1}                &\ctrl{-1}&\gate{H}&\meter
				\\
				&\lstick{\ket{0}}        &\gate{H} &\ctrl{-1}&\ctrl{1} &\gate{e^{-i\alpha X} \,\, \text{or}\,\, e^{-i\alpha Y}}&\qw       &\ctrl{-1} &\qw
				& \ctrl{-1}              &\qw      &\gate{H}&\meter
				\\
				&\lstick{\ket{0}}        &\gate{H} &\qw      &\ctrl{-1}&\qw                  &\qw       &\qw        &\qw
				&\qw                     &\qw      &\rstick{\ket{\psi_{out}}}
			}
		\end{minipage}
		\caption{Circuit for the measurement-based teleportation algorithm with symmetry-preserving or symmetry-breaking perturbations.}
	\end{figure}
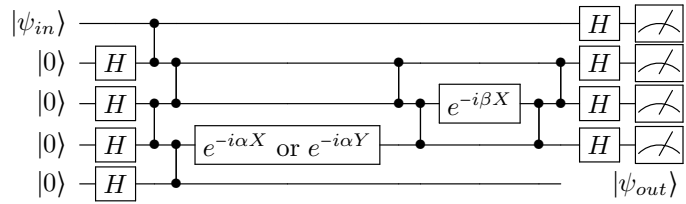
	
\end{widetext}

\end{document}